\documentclass[reprint,twocolumn,amsmath,amssymb,aps,superscriptaddress]{revtex4-2}

\setcitestyle{authoryear,round}
\makeatletter
\AtEndDocument{%
  \immediate\write\@bibdataout{@CONTROL{aip41Control,pages="1",title="0"}}%
  \if@filesw
    \immediate\write\@auxout{\string\citation{aip41Control}}%
  \fi
}
\makeatother
\makeatother
\usepackage{graphicx}
\usepackage{dcolumn}
\usepackage{bm}
\usepackage{graphicx}
\usepackage{amssymb}
\usepackage{epstopdf, epsfig}
\usepackage{yfonts}
\usepackage[colorlinks=true, linkcolor=blue, citecolor=blue, urlcolor=blue]{hyperref}
\usepackage{fontenc}
\usepackage{upgreek}
\usepackage{mathrsfs}
\usepackage{booktabs}
\usepackage{pifont}
\usepackage{amsmath}
\usepackage{bm}
\usepackage{wasysym}
\usepackage{caption}
\usepackage{subcaption}
\usepackage{euscript}
\usepackage{natbib}
\usepackage{lipsum}
\usepackage{xcolor}

\usepackage{orcidlink}

\usepackage{subcaption}
\usepackage{ragged2e}
\DeclareCaptionJustification{justified}{\justifying}
\newcommand{\x}[1]{\textcolor{black}{#1}}

\newcommand{\jfm}[1]{\textcolor{blue}{#1}}

\begin{document}

\title{Attenuation of shoaling extreme wave crests by artificial salt marshes}

\author{Zachary Quek\, \orcidlink{0009-0004-6188-2323}}
\affiliation{School of Civil and Environmental Engineering, Nanyang Technological University, Singapore}

\author{Sahana Mahesh}
\affiliation{School of Civil and Environmental Engineering, Nanyang Technological University, Singapore}

\author{Martin Fronty\,\orcidlink{0009-0007-0030-1902}}
\affiliation{School of Civil and Environmental Engineering, Nanyang Technological University, Singapore}

\author{Seungo Baek\,\orcidlink{0009-0001-1778-7079}}
\affiliation{School of Civil and Environmental Engineering, Nanyang Technological University, Singapore}

\author{Gun Kim\,\orcidlink{0000-0002-2983-5808}}
\email{gun.kim@ntu.edu.sg}
\affiliation{School of Civil and Environmental Engineering, Nanyang Technological University, Singapore}

\author{Saulo Mendes\,\orcidlink{0000-0003-2395-781X}}
\email{saulo.mendes@ntu.edu.sg}
\affiliation{School of Civil and Environmental Engineering, Nanyang Technological University, Singapore}

\begin{abstract}
Laboratory experiments in a wave flume investigated the attenuation of shoaling extreme waves by artificial salt marshes placed immediately before an abrupt depth transition. Increasing vegetation density systematically reduced the excess kurtosis of the surface elevation distribution in the intermediate water depth regime. Nonetheless, attenuation exhibits a short saturation regime at low vegetation densities, whereas at very high densities, excess kurtosis decreases significantly across the gauges. However, kurtosis attenuation becomes ineffective as waves approach shallow-water conditions or as second-order nonlinear effects become significant. Furthermore, the highest experimental densities where the kurtosis is tamed exceed typical natural salt marsh stem densities. Thus, in natural settings, salt marshes are likely to provide only partial mitigation of extreme wave statistics in intermediate and shallow water depths, as the wave field retains some hazardous nonlinearity. 
\end{abstract}

\maketitle

\section{Introduction}\label{sec:intro}

Coastal vegetation is increasingly used as a wave attenuator for storm surges and large wind waves \citep{Tsai2023}, and vegetated foreshores also preserve habitat and retain the capacity to recover following extreme events \citep{lei2019wavedamping}. Generally, this attenuation is governed by canopy-scale hydrodynamics: emergent arrays convert mean kinetic energy into stem-scale turbulence, whereas dense submerged meadows generate a shear layer whose vortices penetrate only a finite depth into the stand \citep{nepf1999drag,nepf2012flow}. While studies have typically focused on the attenuation of bulk wave properties \citep{willemsen2026contribution}, effective coastal protection must also account for extreme-value variables. To the best of our knowledge, no existing study has examined how the attenuative capacity of vegetation influences extreme wave statistics under strongly non-homogeneous conditions. The closest analysis for homogeneous settings in the literature remains the work of \citet{Losada2004} and \citet{henry2013wave}, who re-evaluated the impact of vegetation on the spatial evolution of the root-mean-square wave height ($  H_{\mathrm{rms}}  $) by assuming a Rayleigh distribution of wave heights. Likewise, \citet{chen2012theoretical} have used a joint distribution of wave heights and periods \citep{Higgins1975} to recompute the wave dissipation rate and extended it to broad-banded wave fields. However, the Rayleigh distribution is applicable primarily to narrow-banded, linear irregular waves in deep water \citep{Higgins1980}, where the influence of submerged vegetation is negligible. Using \citeauthor{Dalrymple1984}'s cylinder model for the bulk wave height decay, a few works have attempted to compare the probability density of wave heights collected in field sites with a closed-form modified Rayleigh distribution using a change of variables \citep{jadhav2013probability,lovstedt2010wave}. However, these distributions were limited to data within the bulk of the wave distribution (tails lighter than the Rayleigh distribution) and subject to very mild slope conditions, still far from extremely nonlinear regimes. Nevertheless, \citet{zhao2024parameterisation} have studied the skewness and wave asymmetry over a very mild slope ($\nabla h = 1/20$), and thus limited the wave climate to small excess kurtosis \citep{Mendes2023b,Zhang2024}. 

\begin{figure*}
\centering
\includegraphics[scale=0.4]{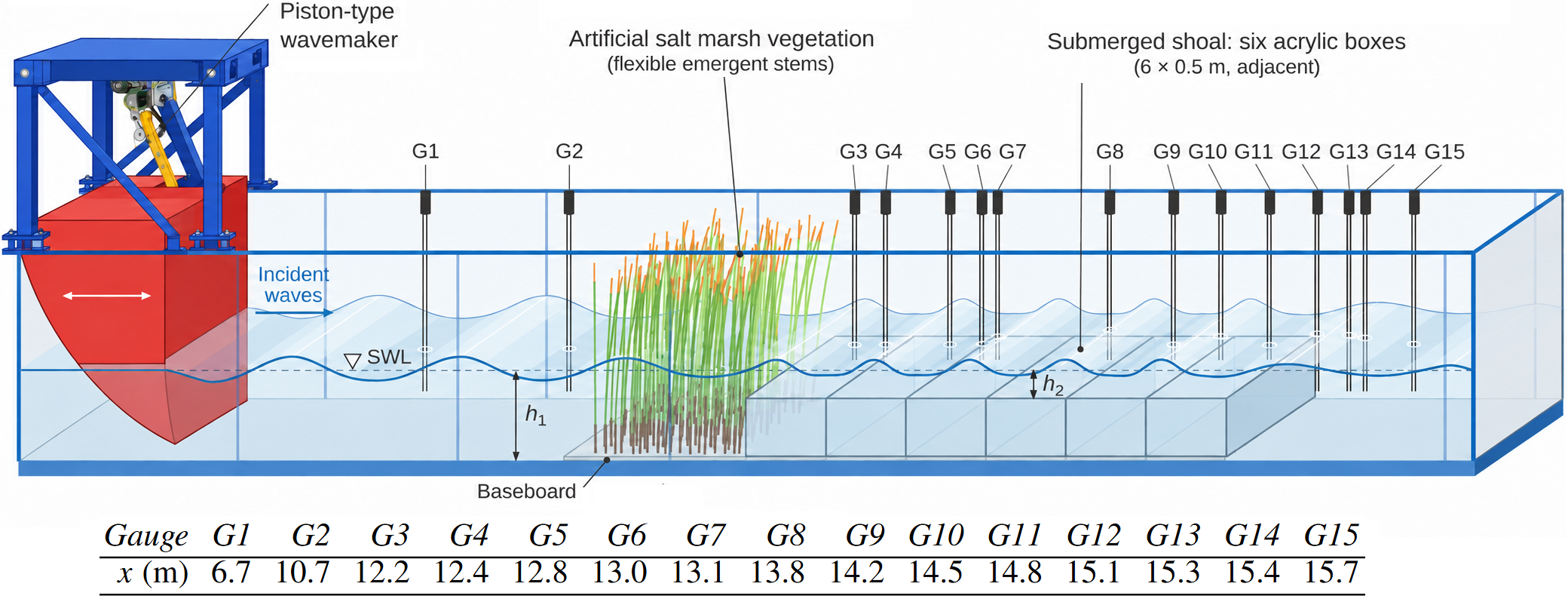}
\caption{Experimental setup of the wave flume (not to scale) with marsh vegetation and acrylic boxes. The schematic illustrates the flume, the wavemaker paddle, and the arrangement of physical test components. At the end of the wave flume, a smooth 1:18 slope extends for 8 m of horizontal length, followed by an absorber. The location of the wave gauges relative to the piston wavemaker is also provided. The distance from the wavemaker to the end of the flume is 32.8 m.}
\label{fig:SketchX}
\end{figure*}
Recent studies have shown that under non-equilibrium conditions, such as waves travelling over relatively steep beaches or steep breakwaters \citep{Chabchoub2019,Trulsen2020,Chabchoub2023,Mendes2024b}, and interacting with non-stationary or non-homogeneous currents \citep{Benoit2023,Mendes2025symmetric,Mendes2026direction}, abrupt nonlinear dynamics emerge and transform the wave-height distribution away from the Rayleigh form into a modified shape that has not been captured by Weibull or standard extreme-value distributions. This behaviour can be explained by physical models that are sensitive to changes in relative water depth and wave steepness, even within the surf zone and despite dissipation from wave breaking \citep{Mendes2026energetics}. A natural expectation grounded in the current literature is that vegetation should attenuate extreme waves. Although there are no studies quantifying the drop in extreme wave statistics as a function of key parameters such as vegetation density, one would expect that a monotonic increase in stem density would ultimately lead to a halt in extreme wave statistics. 

In this work, although restricted to salt marshes, the degree of attenuation of extreme wave statistics by vegetation is measured for the first time. We restrict our study to irregular long-crested waves of small amplitude, covering both linear and weakly nonlinear regimes. To achieve optimal wave attenuation, we use artificial salt marshes in which the leaf (or blade) length exceeds the water depth in all cases. Fixing the physical water depth and leaf length, we examine only variations in dimensionless parameters, such as wave steepness and relative water depth, that are related to the extreme wave formation and activity. For intermediate water waves, we confirm the expectation and observe that a monotonic increase in stem density is met by an almost monotonic decrease in extreme wave statistics, with a distinct saturation regime for low stem density followed by a faster decrease at very high stem densities. Surprisingly, experimental cases with shallower water conditions and the same physical depth atop the step or higher wave steepness show that, even at stem densities far exceeding those found in natural settings, the artificial salt marshes do not mitigate extreme wave hazard.

\section{Experimental Setup and Methodology}\label{ssec:method}

Laboratory studies demonstrate that vegetation situated seaward of coastal dikes and dune faces can attenuate incident waves before they reach these steeper slopes \citep{lovaas2001effect,Baker2022,van2022wave}. The present experiments reproduce this field arrangement by placing a vegetation canopy on a shallow flat followed by an abrupt depth transition, see \jfm{figure} \ref{fig:SketchX}. A vertical step is used in place of a finite slope because the amplification of extreme waves increases with slope steepness and saturates at the steepest transitions, so that a step recovers the same highest amplification as a steep slope \citep{Adcock2021c,Mendes2022,Mendes2023b}. The setup includes two sections of marsh vegetation situated in the central portion of the flume, followed by acrylic blocks installed at the bed level. Fifteen wave gauges are positioned along the flume to monitor water surface elevations, with $h_1$ and $h_2$ representing specific depths of interest for wave analysis. The flume is 35 m long, of which 32.8 m extends downstream of the wavemaker. The final 8 m of the tank has a sloping smooth beach with $\nabla h \approx 1/18$, with 2 m of this length being covered by sponges for reflection mitigation. The six acrylic boxes are adjacent to each other, each measuring 0.5 m in length, spanning a total of 3 m and placed from $x = 11.8$ m to $x+\Delta x = 14.8$ m. Gauges G3-G11 are thus located within the shallower region. Greater wave attenuation is expected under emergent conditions because the plant stems span the entire water column, whereas near-emergent conditions leave the upper water column where orbital velocities are highest, unimpeded by the vegetation \citep{augustin2009laboratory}.
\begin{figure*}
\centering
\includegraphics[scale=0.23]{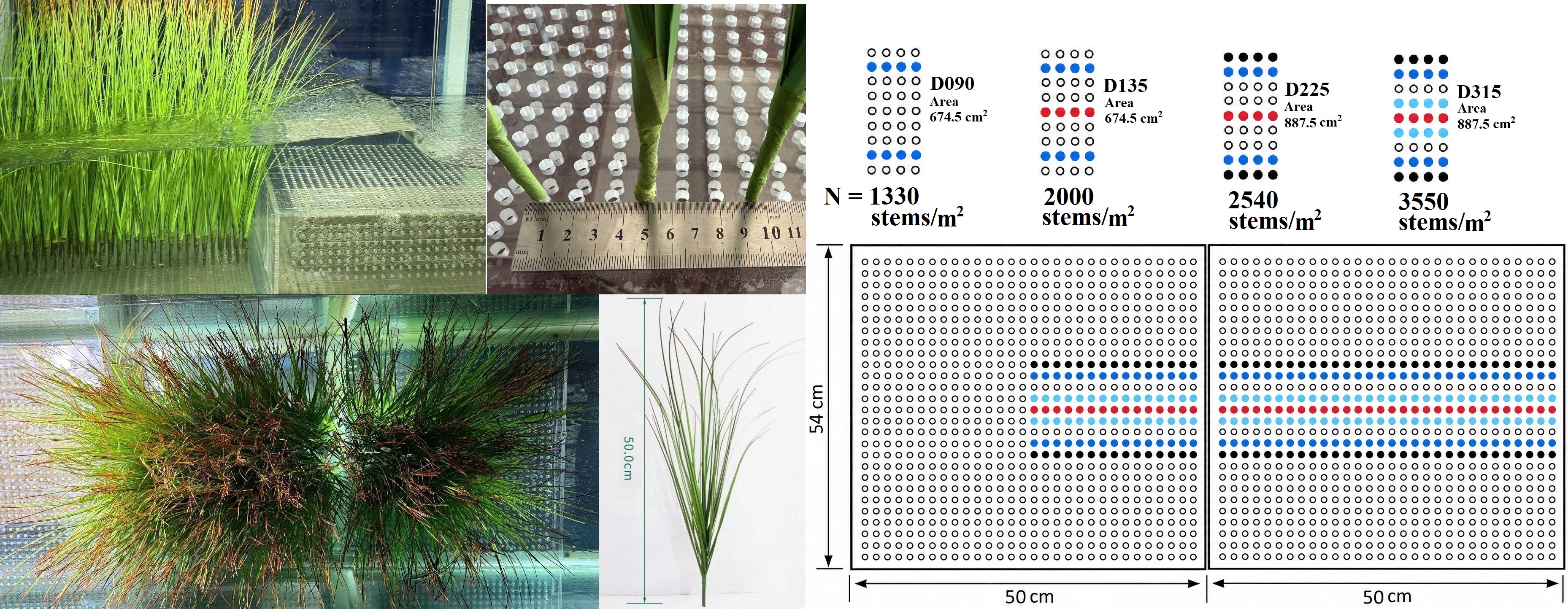}
\caption{Side and top views of the experimental setup (left), and baseboard stem distribution (right). The baseboard has open orifices, whereas the step acrylic box has covered ones.}
\label{fig:SketchX2}
\end{figure*}

To characterize the stiffness of the artificial vegetation, its Young’s modulus ($E$) was estimated from the mass density ($\rho$) and the ultrasonic P-wave velocity ($V_p$). A single plant, comprising a stem and fifteen blades, has a mass of 8.98 g and a volume of 16 $\textrm{cm}^3$ and stem diameter $D = 5$ mm, giving $\rho$ of approximately 560 $\textrm{kg}/\textrm{m}^3$. $V_p$ was measured across the blade thickness (0.2 mm) of four plants in through-transmission mode, using a pair of 20-MHz transducers (V208-RM, Evident, USA) and an ultrasonic measurement system (RAM-5000, RITEC Inc., USA) \citep{BAEK2025139771}, yielding a mean value of 2,095.8 $\pm$ 152.1 m/s (two standard deviations). Assuming isotropic linear elasticity and a Poisson’s ratio 
($\nu$) of 0.3-0.4, $E$ was then obtained from these two quantities through the bulk P-wave relation \citep{barber2004elasticity}. Averaging the estimates over the two bounds of $\nu$ yields $ E = 1.49 \pm 0.22 $ GPa (two standard deviations across plants), with the choice of $\nu$ alone shifting the estimate by approximately $\pm$ 0.34 GPa, where the former reflects plant-to-plant variability and the latter reflects sensitivity to the assumed $\nu$. The artificial vegetation is therefore more compliant and less dense than natural salt marshes, with $E$ ranging from 2 to 10 GPa and a typical value close to 5 GPa \citep{Goseberg2024,liu2025seasonal}, and a typical mass density of $700$ kg/$m^3$ \citep{donatelli2018seagrass}. 

The drag coefficient $C_D$ has been obtained through \citeauthor{Dalrymple1984}'s model for irregular waves \citep{Losada2004}, see also \citet{Gary2026} (the submerged water depth is $h_{vs} = h$):
\begin{widetext}
\begin{equation}
\frac{H_{\mathrm{rms}}}{H_{\mathrm{rms},0}}=\frac{1}{1+\alpha x} \,\, , \,\, \alpha
=\frac{1}{3\sqrt{\pi}}C_D D N ( k_p H_{\mathrm{rms},0} )
\Bigg[\frac{\sinh^3 (k_p h)+3\sinh (k_p h)}{(\sinh (2k_ph)+2k_ph)\sinh (k_ph)} \Bigg]  \, ,
\label{eq:Dalrymple}
\end{equation}
\end{widetext}
with $D$ being the stem diameter, $N$ the stem density per unit area, $k_p$ the peak wavenumber computed from the dispersion relation originating from the peak period $T_p$, and $H_{\mathrm{rms}} = 2\sqrt{2m_0}$ is the root-mean-square wave height (under the narrow-band linear-wave approximation) computed directly from the JONSWAP spectral shape:
\begin{widetext}
\begin{equation}
\label{eq:jonswap}
  S(f)=\frac{\mathcal{E}{g}^2}{\left(2\pi\right)^4}\frac{1}{f^5}\exp{\left[-\frac{5}{4}\left(\frac{f_p}{f}\right)^4\right]}\gamma^{\exp{\left[-\frac{\left(f-f_p\right)^2}{2\left(\delta f_p\right)^2}\right]}} \quad , \quad m_0  = \int_{0}^{\infty} S(f) df  \, ,
\end{equation}
\end{widetext}
where $f_p=1/T_p$ denotes the spectral peak frequency, $\mathcal{E}$ the spectral energy parameter, $\gamma$ the peakedness parameter, and $\delta$ the width parameter, $\delta=0.07$ for $f<f_p$ and $\delta=0.09$ for $f\geq f_p$.

\begin{table*}
  \centering
\begin{tabular}{rrrrrrrrrrrrrrrr}    
\toprule
\emph{Run} & \emph{$\gamma$}  & \emph{$k_{p1}h_{1}$} &  \emph{$k_{p3}h_{3}$}  &  \emph{$k_{p1}H_{s1}$}  & \emph{$k_{p3}H_{s3}$} &   \emph{$H_{s1}$}  &   \emph{$H_{s3}$}  &  \emph{$\lambda_{p1}$}  &  \emph{$\lambda_{p3}$} &  \emph{$h_{1}$} &  \emph{$h_{3}$}  &  \emph{$\bf{C_{D}}$} &   \emph{$\bf{KC}$} &   \emph{$\bf{\alpha}$}  &   \emph{$\alpha_{\mu_4}$}  \\
\cline{1-16}
A & 7 & 0.75 & 0.40 & 0.055 & 0.095 & 0.019 & 0.021 & 2.15 & 1.40 & 0.26 & 0.09 & $|$ \bf{1.70} & \bf{22.7} & \bf{0.15} & 0.75\\
B & 7 & 1.10 & 0.60 & 0.130 & 0.230 & 0.031 & 0.035 & 1.50 & 0.95 & 0.26 & 0.09 & $|$ \bf{1.85} & \bf{29.9} & \bf{0.28} & 0.21\\
C & 7 & 1.45 & 0.70 & 0.170 & 0.250 & 0.030 & 0.032 & 1.12 & 0.80 & 0.26 & 0.09 & $|$ \bf{2.03} & \bf{25.6} & \bf{0.31} & 0.22\\
D & 3 & 1.45 & 0.70 & 0.175 & 0.260 & 0.031 & 0.033 & 1.12 & 0.80 & 0.26 & 0.09 & $|$ \bf{2.15} & \bf{26.3} & \bf{0.34} & 0.23\\
E & 7 & 1.45 & 0.70 & 0.100 & 0.155 & 0.018 & 0.020 & 1.12 & 0.80 & 0.26 & 0.09 & $|$ \bf{2.68} & \bf{16.7} & \bf{0.27} & 1.28\\
F & 7 & 1.45 & 0.70 & 0.195 & 0.275 & 0.035 & 0.035 & 1.12 & 0.80 & 0.26 & 0.09 & $|$ \bf{2.11} & \bf{28.1} & \bf{0.35} & $<$0.01\\
\cline{1-16}\\
\end{tabular}
\caption{Summary of conditions (rounded to the displayed precision) before the step (gauge 1) and at the first gauge located on the step (gauge 3) for each run, with no vegetation. Wave conditions are averaged over ten realizations, each realization containing 1,000 waves. Indices 1 and 3 stand for gauge indices, describing offshore and shallow water conditions. Without loss of generality, we have computed the drag coefficient, the damping coefficients for the rms wave height $\alpha$ and the Keulegan–Carpenter number over a flat bottom (without the step transition) for the intermediate vegetation density $N = 2000$ \textrm{stems}/$\textrm{m}^2$ for the same conditions at gauge G1. Dimensional parameters ($H_{si},h_{i},\lambda_{pi}$) have units of meters or their inverse $\textrm{m}^{-1}\, (\alpha,\alpha_{\mu4})$. The excess kurtosis damping coefficient $\alpha_{\mu4} = [(\mu_{4,0}/\mu_{4}) - 1]/\Delta x$ is computed for the step transition.}
\label{tab:1}
\end{table*}
\begin{figure*}
\centering
\includegraphics[width=0.999\textwidth]{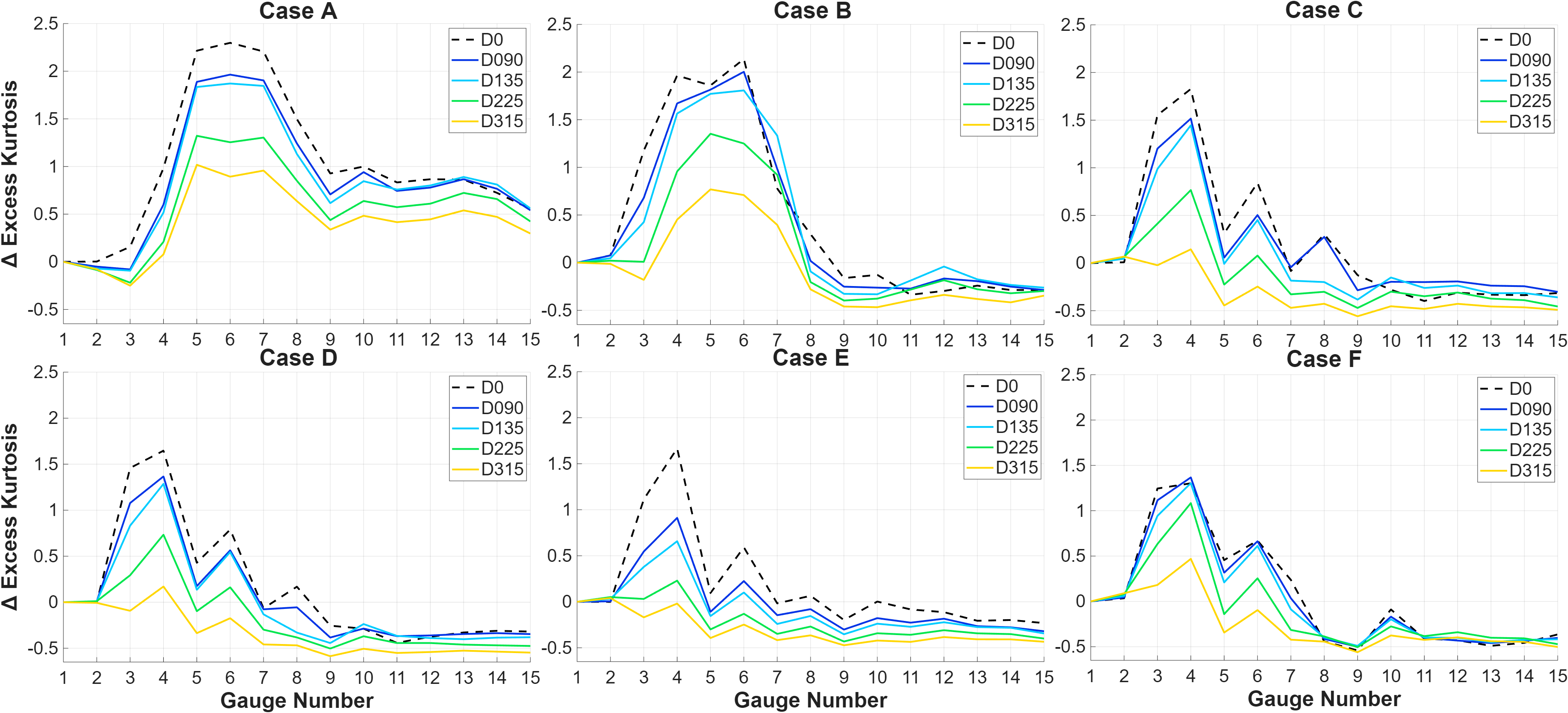}
\caption{Spatial variability of excess kurtosis as a function of vegetation density.}
\label{fig:Kurtosis of 0-2y}
\end{figure*}
Vegetation density is provided in \jfm{figure} \ref{fig:SketchX2}, where, for instance, D090 conveys the total count of 90 stems (with 15 blades per stem), which leads to a vegetation density of $N = 90 / 0.06745 \approx 1330 \, \textrm{stems}/\textrm{m}^2$. While the stem patch measures 71 cm at rest, its horizontal extent can reach 100 cm as the blades deform under wave action. The wave gauges are positioned along the flume width centreline, and the vegetation is distributed densely around the centreline. As observed in \jfm{Table} \ref{tab:1}, and without loss of generality for choosing the particular case of $N = 2000 \, \textrm{stems}/\textrm{m}^2$ (D135), the drag coefficient is consistently found to be in the range $1.7 < C_D <  2.7$. This agrees well with field measurements of salt marsh drag derived from biomechanical properties \citep{Pruter2026}. Additionally, \citeauthor{marsooli2017modeling}'s study showed that most spatially distributed values of the drag coefficient lie in the interval $1< C_D < 2$ over large areas. The peak horizontal orbital velocity in the direction of motion can be estimated using linear theory, and the peak Keulegan–Carpenter number can be computed as $KC = u_{max} T_p/D \approx \pi H_s \coth{(k_p h)} / D $ in linear theory \citep{anderson2014wave,ryan2025influence}, see \jfm{table} \ref{tab:1}. The values of the drag coefficient and KC number follow the trends observed in figures 9b and 10b of \citet{anderson2014wave} for artificial salt marshes. Hence, our experimentation provides similar properties to those of real salt marshes, allowing us to infer how vegetation might tackle extreme waves and obtain a quantitative analysis of their attenuative power.

\begin{figure*}
\centering
\includegraphics[width=0.95\textwidth]{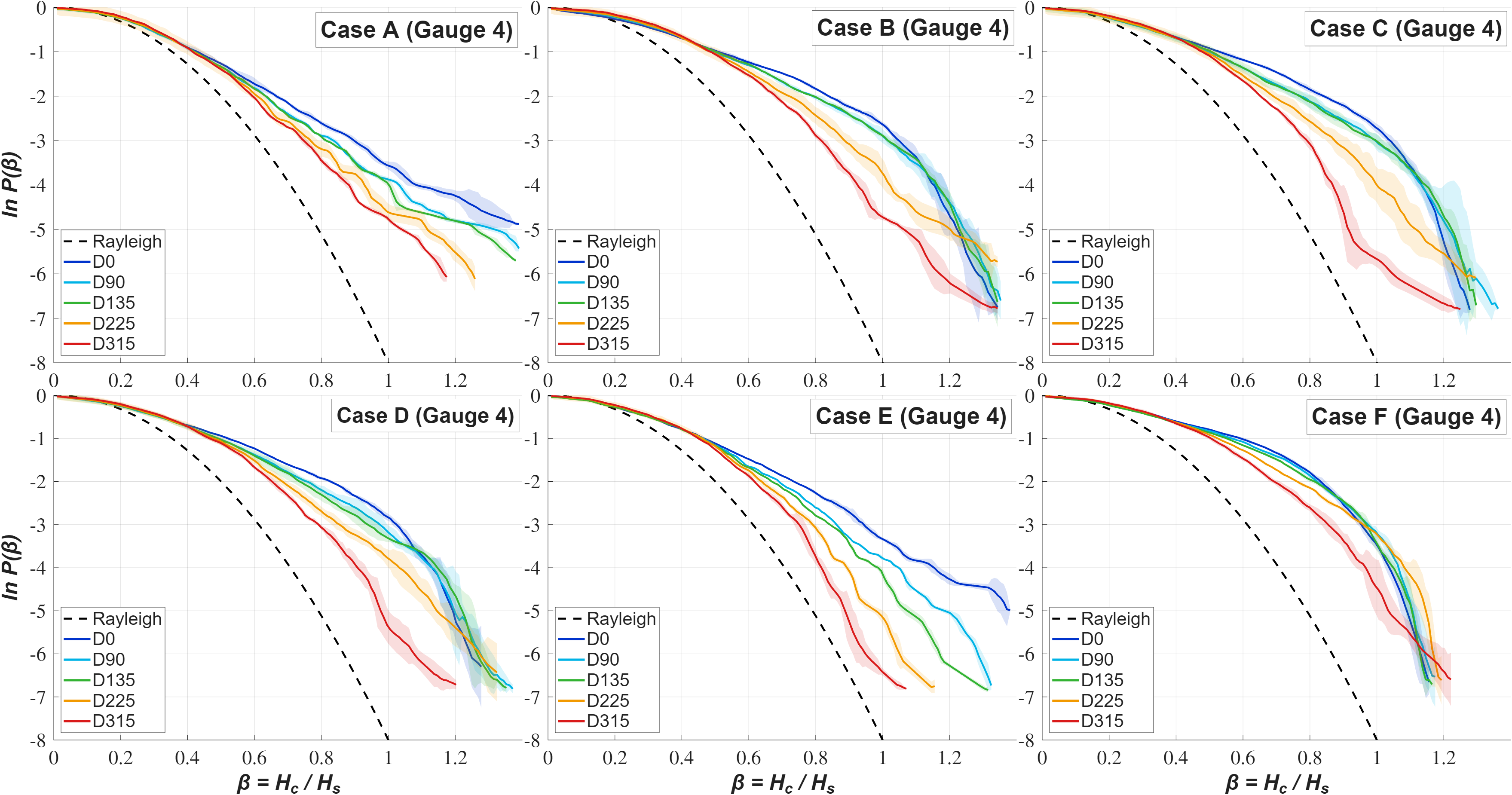}
\caption{Crest exceedance probability reduction due to vegetation density.}
\label{fig:Crest Exceedance Probability all densities}
\end{figure*}
\begin{figure*}
\centering
\includegraphics[width=0.95\textwidth]{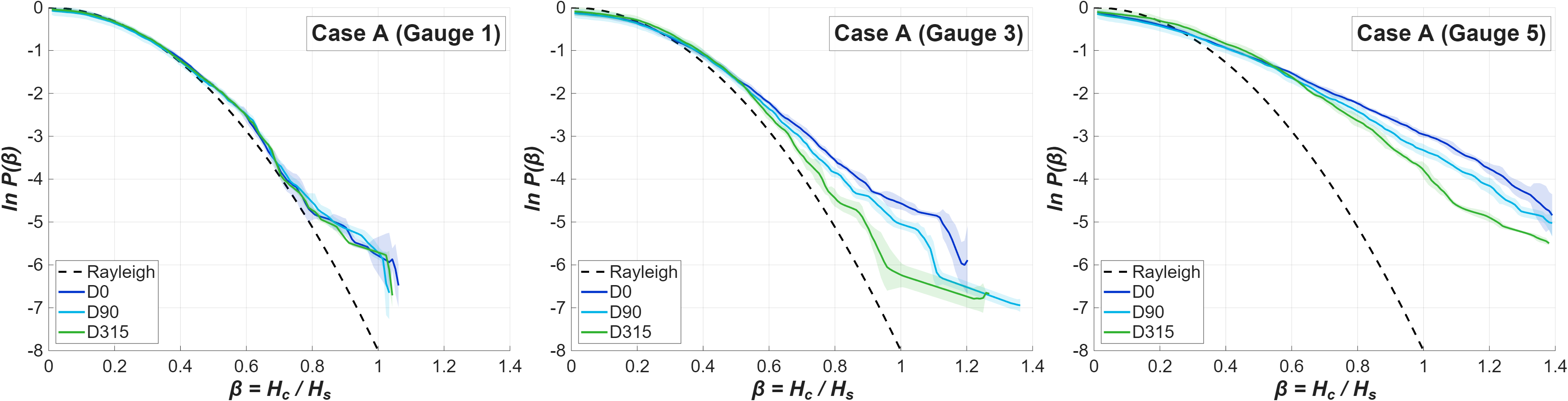}
\caption{Variation in the crest exceedance probability reduction due to vegetation across different gauges.}
\label{fig: 1Fy Crest Exceedance Probability D0,90,315}
\end{figure*}

\section{Results and Discussions}\label{ssec:results} 

In shallower depths, the attenuation caused by submerged vegetation is strongly dependent on the dimensionless parameter $k_p h$. Nonetheless, we do not vary water depth, and the wavenumber takes a leading role. Consequently, the combination of hyperbolic functions approaches a constant lower bound $\alpha \rightarrow (C_D D N H_{\mathrm{rms},0}/3\sqrt{\pi}) \cdot (3/4h)$ at fixed physical water depth $h$ and vegetation properties as we approach the limit $k_p h \rightarrow 0$. We observe this in \jfm{Table} \ref{tab:1}, where the damping coefficient $\alpha$ is weaker in shallow water. The decay in significant wave height thus varies from ~25\% for the largest values of $\alpha$ (cases C, D and F) and 13\% for the smallest one (case A). 

\x{To extend this analysis of mitigation to extreme waves in the same manner, we must choose an integral measure that conveys information thereof. Extreme waves can arise from storms \citep{fedele2010long} and their surges \citep{gao2025extreme}, hurricanes approaching the coast \citep{wang2005extreme}, rogue wave mechanisms \citep{Chabchoub2023}, among other processes. Regardless of their source characteristics, the evaluation of their extreme character is typically carried out by examining the tail of the surface elevation distribution, or the tail of the crest or height exceedance probability. Since probability densities are not integral measures, to complement the threshold-dependent exceedance curves with a single summary measure of non-Gaussianity, we calculate the excess kurtosis of the surface elevation, see \citet{Mori2002b} for a review. \jfm{Figure} \ref{fig:Kurtosis of 0-2y} depicts variations in excess kurtosis relative to the first gauge (offshore) for all cases in \jfm{Table} \ref{tab:1}. We observe that in cases C-F the highest vegetation density ($N = 3550$ \textrm{stems}/$\textrm{m}^2$) mitigates extreme waves by reducing the relative excess kurtosis to below zero, thus, placing the wave climate in a sub-Gaussian regime. From the perspective of coastal protection measures, the artificial salt marshes perform exceptionally at such high densities. However, in cases A and B, we do not see a complete mitigation of extreme seas even with very high vegetation densities. For typical vegetation densities observed in the vast majority of marsh species ($N \lesssim 2000$ \textrm{stems}/$\textrm{m}^2$, blue and cyan curves) \citep{gleason1979effects,tempest2015review}, reduction of the excess kurtosis is minimal in cases A-C-D, whereas no reduction is found in cases B and F. Only in case E, with small wave steepness in intermediate water, do we observe a significant decay in excess kurtosis.}

\begin{figure*}
\centering
\includegraphics[width=\textwidth]{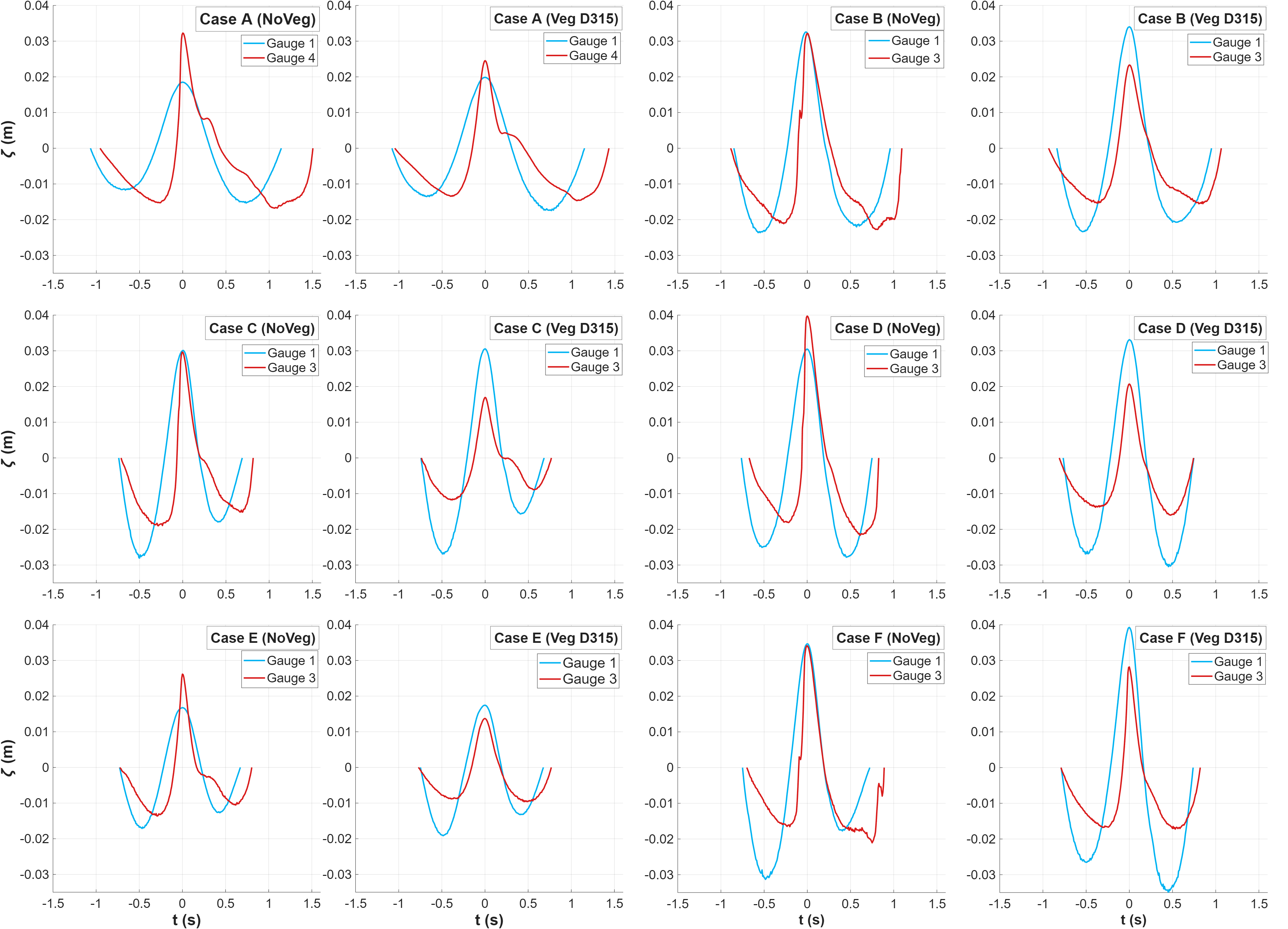}
\caption{Transformation of the largest wave crest at G1 (cyan), recorded at G3 (red) immediately downstream of the vegetation patch, or at G4 for case A, where the kurtosis peak occurs one gauge farther downstream.}
\label{fig:WaveEvol}
\end{figure*}
To compare statistical and bulk measures of wave attenuation by vegetation, we introduce the excess kurtosis damping coefficient, henceforth defined as $\alpha_{\mu4} := [(\mu_{4,0}/\mu_{4}) - 1]/\Delta x$. However, the excess kurtosis will grow even if vegetation is present between gauges G2 and G3, and this definition is applied to compare cases with and without vegetation at G4 (typical peak of kurtosis across most cases), for instance $N= 0$ and $N = 2000$ \textrm{stems}/$\textrm{m}^2$ in \jfm{table} \ref{tab:1}. Thus, $\alpha_{\mu4}$ measures the degree of damping of the kurtosis evolution from gauges G2 to G4. When comparing in \jfm{Table} \ref{tab:1} the variations in the damping coefficient $\alpha$ with the variation in the excess kurtosis $\alpha_{\mu4}$, we see two different trends for the intermediate vegetation density of $N = 2000$ \textrm{stems}/$\textrm{m}^2$. For cases B–D, the damping coefficients of kurtosis and significant wave height increase simultaneously, exhibiting a strong positive Pearson correlation of $r = 0.99$. In contrast, cases A and E–F exhibit a less systematic relationship, resulting in an overall negative correlation $(r = -0.56)$ across all cases. Remarkably, characterised by the steepest waves, case F exhibits no kurtosis decay despite experiencing the strongest attenuation of significant wave height. This contrasting behaviour suggests that wave steepness and shallow-water effects can substantially decouple kurtosis attenuation from wave-height attenuation. Consequently, the relationship between the attenuation of extreme-wave statistics and significant wave height depends strongly on the combined wave–vegetation–structure conditions.

\x{On the other hand, a complementary analysis of crest-height statistics is also required to evaluate the maximum attainable crest height and its frequency of occurrence. Particularly for coastal structures, which cannot follow the crest-to-trough wave motion, maximum extreme crest heights are of paramount importance. In \jfm{figure} \ref{fig:Crest Exceedance Probability all densities}, the exceedance probability of normalized crest heights (relative to the significant wave height) measured at gauge G4 is featured, together with error bands computed with two standard deviations among the ten realizations of 1,000 waves each. Gauge G4 is chosen due to being the most common location of maximum kurtosis among all cases. We compare the experimental data with the Rayleigh distribution (dashed) \citep{Higgins1952}, and thus the more extreme the waves are, the larger the gap between the experimental curve and the dashed one. For coastal protection measures, exceptional mitigation can be attributed to a reduction in this gap, ideally bringing the exceedance below the Rayleigh curve. Cases A and E show a stratified crest distribution for all $\beta = H_c/H_s$, demonstrating a monotonic reduction of the gap to the Rayleigh curve as vegetation density increases. For cases B-C-D the monotonic reduction also holds, except at the tail ($\beta > 1.2$). In agreement with the lack of kurtosis decay, case F has no reduction in the gap between observed exceedance crest probability and the Rayleigh distribution. Notably, as observed for the excess kurtosis, the reductions in this gap are very small at vegetation densities typical of natural settings, except for case E.} Although we have chosen gauge G4 for this exposition, this choice does not restrict the generality of our observations: \jfm{figure} \ref{fig: 1Fy Crest Exceedance Probability D0,90,315} shows that, except for the offshore gauges, atop the acrylic boxes following the ADT, we observe the very same features of gauge G4 in other gauges.

Finally, \jfm{figure} \ref{fig:WaveEvol} selects the highest wave crest at gauge G1 and tracks it along the path towards the vegetation patch and subsequently the ADT, translating the risk of extreme waves from a probability framework to a strictly coastal protection perspective. We observe the same trend as before with kurtosis or exceedance probability, now for the tallest wave crests: in cases C-E the extreme crest is significantly reduced (40-50\%) by the artificial salt marshes, whereas for cases A-B and F we see marginal ($\lesssim$ 20\%) mitigation even with a very high density of vegetation ($N = 3550$ \textrm{stems}/$\textrm{m}^2$, above most reported natural salt-marsh densities), with no appreciable mitigation found for densities compatible with natural conditions ($N = 1330$ \textrm{stems}/$\textrm{m}^2$) in the corresponding data.

\section{Conclusions \label{sec:conclusion}}

In this work, we evaluated the extent to which vegetation density decreases the extreme nonlinear behaviour. Such attenuation proved to be weaker than anticipated, and in some cases far smaller than the mitigation of the significant wave height. Even for very high vegetation density compatible with the maximum values observed in natural settings, the system is still nonlinear and hazardous despite attenuation by salt marshes. Only when salt marsh vegetation density exceeds that typically found in natural settings does the kurtosis approach Gaussian levels, and even then, not in all cases. Thus, we find that salt marshes have a reduced impact on mitigating extreme wave crests before an ADT. This suggests that under non-homogeneous or non-stationary conditions, including those associated with tidal currents, undertow, sea-level rise, storm surges or erosion, vegetation may attenuate extreme waves to a much lesser degree than otherwise over stationary and/or homogeneous conditions. 

Furthermore, our results highlight that there are relative water depth regimes where the attenuative power of vegetation against extreme wave statistics is optimized. As such, the prospect of hybrid structures (combining vegetation cover with grey structures, such as breakwaters or seawalls) is likely to provide optimal results in intermediate water rather than in shallow water. Our results thus suggest that, as in other interactions between irregular waves and obstacles or flows, the decay of ordinary waves by vegetation (i.e., significant wave height) is not a reliable proxy for a mirrored behaviour in extreme wave maximum or frequency.

Beyond mitigating wave heights, coastal vegetation plays a broader role in ecohydraulics, influencing wave-induced loads, flow kinematics, currents, sediment transport and other related processes. In terms of cumulative wave loads, the concern is not only the maximum load but also the frequency of extreme loading events. Future work should therefore assess how cumulative extreme loads are modified by vegetation, as well as potential trade-offs between high vegetation density over short distances and lower density over longer distances. Further investigations should examine the influence of wave breaking on extreme-wave attenuation and the role of vegetation in slope stability under strongly non-homogeneous conditions, such as the abrupt depth transitions considered in this study.

\section{Acknowledgements} S. M. acknowledges support from the NTU start-up funds under grant 025629-00001.

\bibliography{bibliography}

\end{document}